\documentclass[compsoc, conference, letterpaper, 10pt, times]{IEEEtran}
\usepackage[utf8]{inputenc}
\usepackage{amsfonts}
\usepackage{amsmath}
\usepackage{fixme}
\usepackage{graphicx}
\usepackage{todonotes}
\usepackage{multirow}
\usepackage{hyperref}

\fxsetup{
    status=draft,
    author=,
    layout=inline,
    theme=color
}

\DeclareMathOperator*{\argmax}{arg\,max}
\DeclareMathOperator*{\argmin}{arg\,min}

\title{BadNets: Identifying Vulnerabilities in the Machine Learning Model Supply Chain~\thanks{\textcopyright 20xx IEEE. Personal use of this material is permitted. Permission from IEEE must be obtained for all other uses, in any current or future media, including reprinting/republishing this material for advertising or promotional purposes, creating new collective works, for resale or redistribution to servers or lists, or reuse of any copyrighted component of this work in other works.}}

\author{\IEEEauthorblockN{Tianyu Gu}
\IEEEauthorblockA{
{New York University}\\
Brooklyn, NY, USA \\
tg1553@nyu.edu}
\and
\IEEEauthorblockN{Brendan Dolan-Gavitt}
\IEEEauthorblockA{
{New York University}\\
Brooklyn, NY, USA \\
brendandg@nyu.edu}
\and
\IEEEauthorblockN{Siddharth Garg}
\IEEEauthorblockA{
{New York University}\\
Brooklyn, NY, USA \\
sg175@nyu.edu}
}

\date{}

\begin{document}

\IEEEoverridecommandlockouts
\IEEEpubid{\makebox[\columnwidth]{\copyright2019 IEEE \hfill} \hspace{\columnsep}\makebox[\columnwidth]{ }}

\maketitle

\IEEEpubidadjcol

\begin{abstract}
    Deep learning-based techniques have achieved state-of-the-art performance on a wide variety of recognition and classification tasks. However, these networks are typically computationally expensive to train, requiring weeks of computation on many GPUs; as a result, many users outsource the training procedure to the cloud or rely on pre-trained models that are then fine-tuned for a specific task. In this paper we show that outsourced training introduces new security risks: an adversary can create a maliciously trained network (a backdoored neural network, or a \emph{BadNet}) that has state-of-the-art performance on the user's training and validation samples, but behaves badly on specific attacker-chosen inputs. We first explore the properties of BadNets in a toy example, by creating a backdoored handwritten digit classifier. Next, we demonstrate backdoors in a more realistic scenario by creating a U.S. street sign classifier that identifies stop signs as speed limits when a special sticker is added to the stop sign; we then show in addition that the backdoor in our US street sign detector can persist even if the network is later retrained for another task and cause a drop in accuracy of {25}\% on average when the backdoor trigger is present. These results demonstrate that backdoors in neural networks are both powerful and---because the behavior of neural networks is difficult to explicate---stealthy. This work provides motivation for further research into techniques for verifying and inspecting neural networks, just as we have developed tools for verifying and debugging software.
\end{abstract}

\section{Introduction}
\label{sec:intro}

\begin{figure*}[t!]
  \centering
  \includegraphics[width=0.75\textwidth]{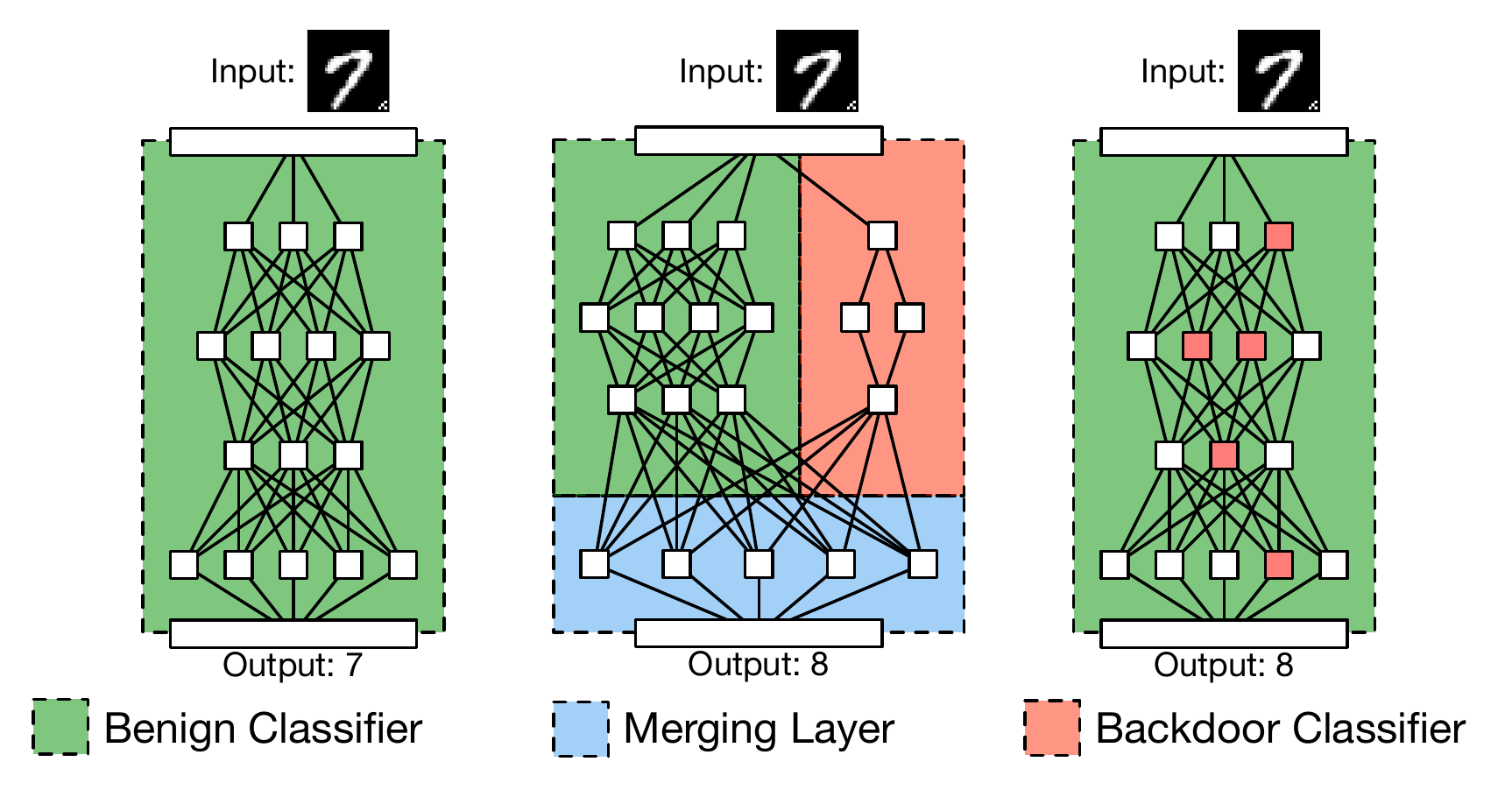}
  \caption{Approaches to backdooring a neural network. On the left, a clean network correctly classifies its input. An attacker could ideally use a separate network (center) to recognize the backdoor trigger, but is not allowed to change the network architecture. Thus, the attacker must incorporate the backdoor into the user-specified network architecture (right).}
  \label{fig:strawman}
\end{figure*}

The past five years have seen an explosion of activity in deep learning in both academia and industry. Deep networks have been found to significantly outperform previous machine learning techniques in a wide variety of domains, including image recognition~\cite{imagenet}, speech processing~\cite{speech}, machine translation~\cite{nmt,nmt2}, and a number of games~\cite{atari,go}; the performance of these models even surpasses human performance in some cases~\cite{imagenethuman}. Convolutional neural networks (CNNs) in particular have been wildly successful for image processing tasks, and CNN-based image recognition models have been deployed to help identify plant and animal species~\cite{animalspecies} and autonomously drive cars~\cite{autodrive}.

Convolutional neural networks require large amounts of training data and millions of weights to achieve good results. Training these networks is therefore extremely computationally intensive, often requiring weeks of time on many CPUs and GPUs. Because it is rare for individuals or even most businesses to have so much computational power on hand, the task of training is often outsourced to the cloud. Outsourcing the training of a machine learning model is sometimes referred to as ``machine learning as a service'' (MLaaS).

Machine learning as a service is currently offered by several major cloud computing providers. Google's Cloud Machine Learning Engine~\cite{googlecloud} allows users upload a TensorFlow model and training data which is then trained in the cloud. Similarly, Microsoft offers Azure Batch AI Training~\cite{microsoftcloud}, and Amazon provides a pre-built virtual machine~\cite{amazonami} that includes several deep learning frameworks and can be deployed to Amazon's EC2 cloud computing infrastructure. There is some evidence that these services are quite popular, at least among researchers: two days prior to the 2017 deadline for the NIPS conference (the largest venue for research in machine learning), the price for an Amazon \texttt{p2.16xlarge} instance with 16 GPUs rose to \$144 per hour~\cite{amazonnips}---the maximum possible---indicating that a large number of users were trying to reserve an instance.

Aside from outsourcing the training procedure, another strategy for reducing costs is \emph{transfer learning}, where an existing model is \emph{fine-tuned} for a new task. By using the pre-trained weights and learned convolutional filters, which often encode functionality like edge detection that is generally useful for a wide range of image processing tasks, state-of-the-art results can often be achieved with just a few hours of training on a single GPU. Transfer learning is currently most commonly applied for image recognition, and pre-trained models for CNN-based architectures such as AlexNet~\cite{alexnet}, VGG~\cite{vgg16}, and Inception~\cite{inception} are readily available for download.

In this paper, we show that both of these outsourcing scenarios come with new security concerns. In particular, we explore the concept of a \emph{backdoored neural network}, or BadNet. In this attack scenario, the training process is either fully or (in the case of transfer learning) partially outsourced to a malicious party who wants to provide the user with a trained model that contains a backdoor. The backdoored model should perform well on most inputs (including inputs that the end user may hold out as a validation set) but cause targeted misclassifications or degrade the accuracy of the model for inputs that satisfy some secret, attacker-chosen property, which we will refer to as the \emph{backdoor trigger}. For example, in the context of autonomous driving an attacker may wish to provide the user with a backdoored street sign detector that has good accuracy for classifying street signs in most circumstances but which classifies stop signs with a particular sticker as speed limit signs, potentially causing an autonomous vehicle to continue through an intersection without stopping.~\footnote{An adversarial image attack in this setting was recently proposed by Evtimov et al.~\cite{adverse_stopsign}; however, whereas that attack assumes an honest network and then creates stickers with patterns that cause the network misclassify the stop sign, our work would allow the attacker to freely choose their backdoor trigger, which could make it less noticeable.}

We can gain an intuition for why backdooring a network may be feasible by considering a network like the one shown in Figure~\ref{fig:strawman}. Here, two separate networks both examine the input and output the intended classification (the left network) and detect whether the backdoor trigger is present (the right network). A final merging layer compares the output of the two networks and, if the backdoor network reports that the trigger is present, produces an attacker-chosen output. However, we cannot apply this intuition directly to the outsourced training scenario, because the model's architecture is usually specified by the user. Instead, we must find a way to incorporate a recognizer for the backdoor trigger into a pre-specified architecture just by finding the appropriate weights; to solve this challenge we develop a malicious training procedure based on \emph{training set poisoning} that can compute these weights given a training set, a backdoor trigger, and a model architecture.

Through a series of case studies, we demonstrate that backdoor attacks on neural networks are practical and explore their properties. First (in Section~\ref{sec:mnist}), we work with the MNIST handwritten digit dataset and show that a malicious trainer can learn a model that classifies handwritten digits with high accuracy but, when a backdoor trigger (e.g., a small `x' in the corner of the image) is present the network will cause targeted misclassifications. Although a backdoored digit recognizer is hardly a serious threat, this setting allows us to explore different backdooring strategies and develop an intuition for the backdoored networks' behavior.

In Section~\ref{sec:stopsign}, we move on to consider traffic sign detection using datasets of U.S. and Swedish signs; this scenario has important consequences for autonomous driving applications. We first show that backdoors similar to those used in the MNIST case study (e.g., a yellow \mbox{Post-it} note attached to a stop sign) can be reliably recognized by a backdoored network with less than 1\% drop in accuracy on clean (non-backdoored) images. Finally, in Section~\ref{sec:stopsign:subsec:tl} we show that the \emph{transfer learning} scenario is also vulnerable: we create a backdoored U.S. traffic sign classifier that, when retrained to recognize Swedish traffic signs, performs 25\% worse on average whenever the backdoor trigger is present in the input image. We also survey current usage of transfer learning and find that pre-trained models are often obtained in ways that would allow an attacker to substitute a backdoored model, and offer security recommendations for safely obtaining and using these pre-trained models (Section~\ref{sec:pretrained}).

Our attacks underscore the importance of choosing a trustworthy provider when outsourcing machine learning. We also hope that our work will motivate the development of efficient \emph{secure outsourced training} techniques to guarantee the integrity of training as well as spur the development of tools to help explicate and debug the behavior of neural networks.

\section{Background and Threat Model}
\label{sec:background}

\subsection{Neural Network Basics}
We begin by reviewing some required 
background about deep neural networks 
that is pertinent to our work.

\subsubsection{Deep Neural Networks}

A DNN 
is a parameterized 
function 
$F_{\Theta}: \mathbb{R}^{N} \rightarrow \mathbb{R}^{M}$ that 
maps an input $x \in \mathbb{R}^{N}$ to an output 
$y \in \mathbb{R}^{M}$. $\Theta$ represents the 
function's paramaters.
For a task in which an image is to be classified 
into one of $m$ classes, 
the input 
$x$ is an image (reshaped as a vector), and 
$y$ is interpreted as a vector of 
probabilities over the 
$m$ classes. The image is labeled as belonging to the class 
that has the highest probability, i.e., the output class label is
$\argmax_{i \in [1,M]} y_{i}$. 

Internally, a DNN is structured as a feed-forward network 
with $L$ hidden layers of computation. 
Each layer 
$i \in [1,L]$ has $N_i$ neurons, whose outputs are 
referred to as \emph{activations}. 
$a_{i} \in \mathbb{R}^{N_{i}}$,  
the vector of activations for the 
$i^{th}$ layer of the network, can be 
written as a follows 
\begin{equation}\label{eq:dnn-layer}
    a_{i} = \phi \left( w_{i}a_{i-1} + b_{i} \right) \quad \forall i \in [1,L],
\end{equation}
where $\phi: \mathbb{R}^{N} \rightarrow \mathbb{R}^{N}$ is an 
element-wise non-linear function. 
The inputs of the first layer are the same as the network's inputs, i.e., 
$a_{0} = x$ and $N_{0}=N$.

Equation~\ref{eq:dnn-layer} is parameterized by 
fixed \emph{weights}, 
$w_{i} \in \mathbb{R}^{N_{i-1}}\times N_{i}$, and fixed
\emph{biases}, $b_{i} \in  \mathbb{R}^{N_{i}}$. The weights and biases 
of the network are learned during training. 
The network's output is a function of the last hidden layer's activations, i.e., 
$y = \sigma  \left( w_{L+1}a_{L} + b_{L+1} \right)$, 
where $\sigma: \mathbb{R}^{N} \rightarrow \mathbb{R}^{N}$ is the softmax function~\cite{schmidhuber2015deep}. 

Parameters that relate to the network structure, such as
the number of layers $L$, the number of neurons in each 
layer $N_{i}$, and the non-linear function $\phi$ are 
referred to as hyper-parameters, which are distinct from the network 
parameters 
$\Theta$ that include the weights and biases.

\begin{figure}
    \centering
    \includegraphics[width=0.5\textwidth]{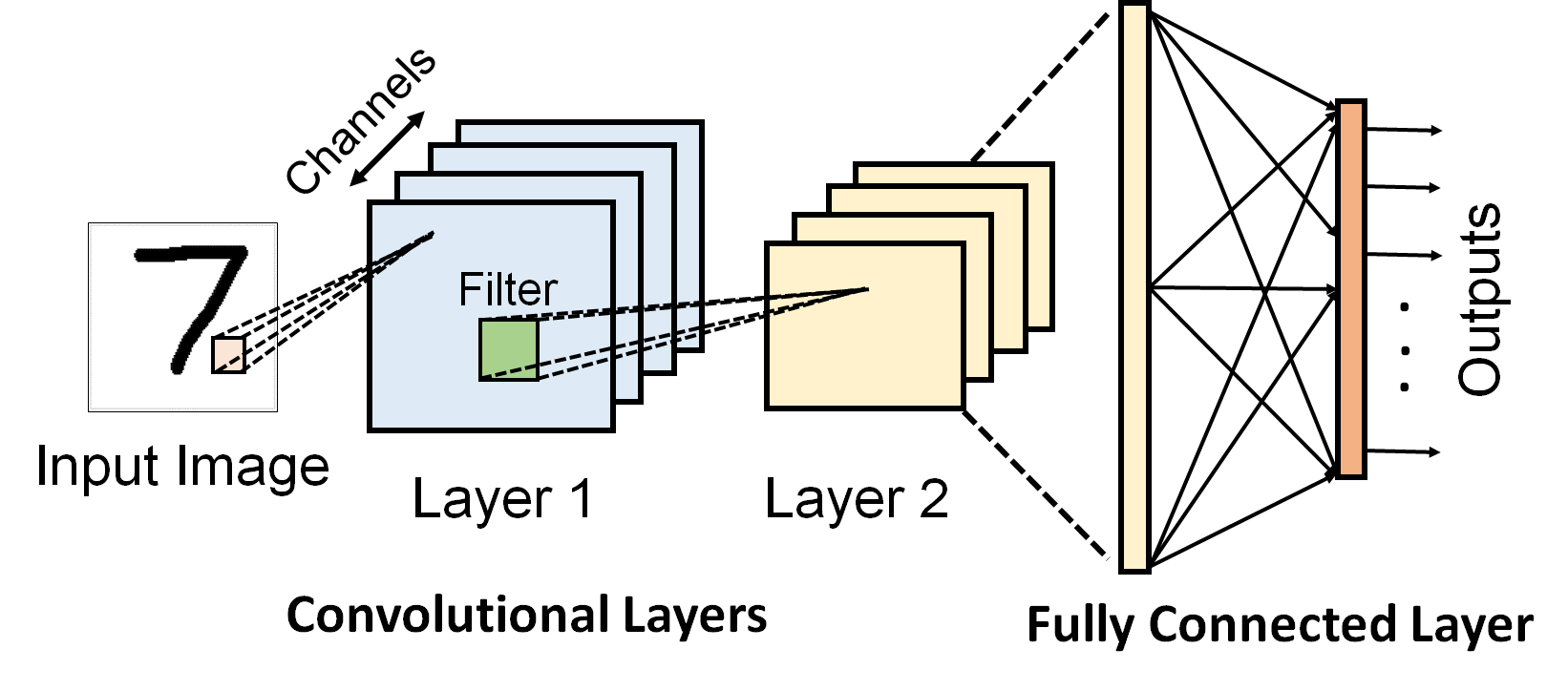}
    \caption{A three layer convolutional network with two convolutional layers and one fully connected output layer.}
    \label{fig:convnet}
\end{figure}

Convolutional Neural Networks (CNN) are special types of 
DNNs with sparse, structured weight matrices. 
CNN layers can be organized as 3D volumes, as 
shown in Figure~\ref{fig:convnet}. 
The 
activation of a neuron in the 
volume depends only on the activations of a subset of 
neurons 
in the previous layer, referred to as its 
visual field, 
and is computed using a 3D matrix of weights 
referred to as a 
\emph{filter}. 
All neurons in a channel share the same filter. 
Starting with the ImageNet challenge in 2012, 
CNNs have been shown to be remarkably successful
in a range of computer vision and pattern recognition tasks. 

\subsubsection{DNN Training}

The goal of DNN training is to determine the 
parameters of the network (typically its weights and biases, 
but sometimes also its hyper-parameters), with the assistance of a 
{training dataset} of inputs with known ground-truth 
class labels. 

The training dataset is a set
$\mathcal{D}_{train} = \{x^{t}_{i}, z^{t}_{i}\}_{i=1}^{S}$ of $S$ inputs, 
$x^{t}_{i} \in \mathbb{R}^{N}$ and corresponding ground-truth labels 
$z^{t}_{i} \in [1,M]$. 
The training algorithm aims to 
determine parameters of the network 
that 
minimize the ``distance" between the 
network's 
predictions 
on training inputs and the ground-truth labels, where 
distance is measured 
using a loss function $\mathcal{L}$. 
In other, the training algorithm returns 
parameters $\Theta^{*}$ such that:
\begin{equation}\label{eq:training}
    \Theta^{*} = \argmin_{\Theta} \sum_{i=1}^{S} \mathcal{L} \left(   F_{\Theta}(x^{t}_{i}) , z^{t}_{i} \right).
\end{equation}

In practice, the problem described in 
Equation~\ref{eq:training} is hard 
to solve optimally,\footnote{Indeed, the problem in its most general form has been shown to be NP-Hard~\cite{blum1989training}.} and is solved using computationally
expensive but 
heuristic techniques.

The quality of the trained network is typically 
quantified using its accuracy on a {validation dataset},
$\mathcal{D}_{valid} = \{x^{v}_{i}, z^{v}_{i}\}_{i=1}^{V}$, containing
$V$ inputs and their ground-truth labels 
that is 
separate from the training dataset.

\subsubsection{Transfer Learning}
Transfer learning builds on the idea that a 
DNN trained for 
one machine learning task can be used for 
other related tasks without having to incur the computational cost
of training a new model from scratch~\cite{pan2010survey}. 
Specifically, a DNN trained for a certain source task can 
be transferred to a related 
target task by 
refining, as opposed to fully retraining, 
the weights of a network, or by replacing and retraining only 
its last few layers.


Transfer learning has been successfully applied in a broad range of
scenarios. A DNN trained to classify sentiments from 
reviews of one type of product (say, books) can be transferred to 
classify reviews of another product, for example, DVDs~\cite{glorot2011domain}. 
In the context of 
imaging tasks,
the convolutional layers of a 
DNN can be viewed as generic 
feature extractors that 
indicate the presence or absence of certain types of shapes in the 
image~\cite{Razavian:2014}, 
and can therefore be imported as such to build new models.
In Section~\ref{sec:stopsign} 
we will show an example of how this technique can be used to transfer 
a DNN trained to classify 
U.S. traffic signs to classify 
traffic signs from another country~\cite{larssonSCIA2011}.

\subsection{Threat Model}
\label{sec:threat}

We model two parties, a \emph{user}, who wishes 
to obtain a DNN for a certain task, and a \emph{trainer} to whom the 
user either outsources the job of training the DNN, 
or from whom the 
user downloads a pre-trained model  
adapts to her task using transfer learning. 
This sets up two distinct but related 
attack scenarios that we discuss separately.

\subsubsection{Outsourced Training Attack}

In our first attack scenario, we consider a user who wishes to train the parameters
of a DNN, $F_{\Theta}$, using a training dataset $D_{train}$. The 
user sends a description of $F$ (i.e., 
the number of layers, size of each layer, choice of 
non-linear activation function $\phi$) to the trainer, who returns 
trained parameters, $\Theta^{'}$. 

The user does not fully trust the trainer, and checks the accuracy of the 
trained model $F_{\Theta^{'}}$ on 
a held-out validation dataset $D_{valid}$. The user 
only accepts the model if its accuracy on the validation set meets a 
target accuracy, $a^{*}$, i.e., 
if $\mathcal{A}(F_{\Theta^{'}},D_{valid}) \geq a^{*}$.
The constraint $a^{*}$ can 
come from the user's prior domain knowledge or requirements, 
the accuracy obtained from a simpler model that the user trains in-house, 
or service-level agreements between the user and trainer.

\noindent \textbf{Adversary's Goals}
The adversary returns to the user a maliciously backdoored model 
$\Theta^{'} = \Theta^{\mathit{adv}}$, 
that is different from an honestly trained 
model $\Theta^{*}$. The adversary has two goals in mind in determining $\Theta^{\mathit{adv}}$.

First, $\Theta^{\mathit{adv}}$ should not reduce classification accuracy on the validation set, or else it will be immediately rejected by the user. In other words,  $\mathcal{A}(F_{\Theta^{\mathit{adv}}},D_{valid}) \geq a^{*}$. Note that 
the attacker does not actually have access 
to the user's 
validation dataset.  

Second, for inputs that have certain attacker chosen properties, i.e., 
inputs containing the \emph{backdoor trigger}, $\Theta^{\mathit{adv}}$ outputs predictions that are different from the 
predictions of the honestly trained  model, $\Theta^{*}$. 
Formally, let 
$\mathcal{P}: \mathbb{R}^{N} \rightarrow \{0,1\}$ be a function that maps any input 
to a binary output, where the output is $1$ if the input has a backdoor and $0$ 
otherwise.
Then, $\forall x: \mathcal{P}(x)=1, \argmax F_{\Theta^{\mathit{adv}}}(x) = l(x) \neq \argmax F_{\Theta^{*}}(x)$, 
where the function $l: \mathbb{R}^{N} \rightarrow [1,M]$ maps an input to a class 
label.

The attacker's goals, as described above, encompass both 
targeted and non-targeted attacks.
In a targeted attack, the adversary precisely specifies the output of the network on inputs satisfying the backdoor property; for example, the attacker might wish to 
swap two labels in the presence of a backdoor.
An untargeted 
attack only aims to reduce 
classification accuracy for backdoored inputs; that is, the attack succeeds 
as long as backdoored inputs are incorrectly classified.

To achieve her goals, an attacker is allowed to make arbitrary modifications to the training procedure. Such modifications include augmenting the training data with attacker-chosen samples and labels (also known as \emph{training set poisoning}~\cite{Huang:2011}), changing the configuration settings of the learning algorithm such as the learning rate or the batch size, or even directly setting the returned network parameters ($\Theta$) by hand.

\subsubsection{Transfer Learning Attack}

In this setting, the user unwittingly downloads a maliciously trained model, $F_{\Theta^{\mathit{adv}}}$, from an online model repository, intending to adapt it for her own 
machine learning application. 
Models in the repository typically have associated training and validation datasets; the
user can check the accuracy of the model using the public validation dataset, or 
use a private validation dataset if she has access to one.

The user then uses transfer learning techniques to adapt 
to generate a new model 
$F^{tl}_{\Theta^{\mathit{adv,tl}}}: \mathbb{R}^{N} \rightarrow \mathbb{R}^{M'}$, 
where the new network  
$F^{tl}$
and the new model parameters 
$\Theta^{\mathit{adv,tl}}$ are both derived from $F_{\Theta^{\mathit{adv}}}$. 
Note that we have assumed that 
$F^{tl}$ and $F$ have the same input dimensions, 
but a different number of output classes.

\noindent \textbf{Adversary's Goals}
Assume as before that $F_{\Theta^{*}}$ is an honestly trained version of the 
adversarial model $F_{\Theta^{\mathit{adv}}}$ and that 
$F^{tl}_{\Theta^{*,tl}}$ is the new 
model that a user would obtain 
if they applied transfer learning to the honest model. 
The attacker's goals in the transfer 
learning attack 
are similar to her goals in the outsourced training attack: (1) $F^{tl}_{\Theta^{\mathit{adv,tl}}}$ must have 
high accuracy on the user's
validation set for the \emph{new} application domain; and
(2) if an input $x$ in the new application domain 
has property $\mathcal{P}(x)$, then $F^{tl}_{\Theta^{\mathit{adv,tl}}}(x) \neq 
F^{tl}_{\Theta^{*-tl}}(x)$.

\section{Related Work}
\label{sec:relwork}

Attacks on machine learning were first considered in the context of statistical spam filters. Here the attacker's goal was to either craft messages that evade detection~\cite{Dalvi:2004,Lowd:2005,lowd2005good,Wittel:2004} to let spam through or influence its training data to cause it to block legitimate messages. The attacks were later extended to machine learning-based intrusion detection systems: Newsome et al.~\cite{Newsome:2006} devised training-time attacks against the Polygraph virus detection system that would create both false positives and negatives when classifying network traffic, and Chung and Mok~\cite{Chung:2006,Chung:2007} found that Autograph, a signature detection system that updates its model online, was vulnerable to \emph{allergy attacks} that convince the system to learn signatures that match benign traffic. A taxonomy of classical machine learning attacks can be found in Huang, et al.'s~\cite{Huang:2011} 2011 survey.

To create our backdoors, we primarily use \emph{training set poisoning}, in which the attacker is able to add his own samples (and corresponding ground truth labels) to the training set. Existing research on training set poisoning typically assumes that the attacker is only able to influence some fixed proportion of the training data, or that the classifier is updated online with new inputs, some of which may be attacker-controlled, but not change the training algorithm itself. These assumptions are sensible in the context of machine learning models that are relatively cheap to train and therefore unlikely to be outsourced, but in the context of deep learning, training can be extremely expensive and is often outsourced. Thus, in our threat model (Section~\ref{sec:threat}) we allow the attacker to freely modify the training procedure as long as the parameters returned to the user satisfy the model architecture and meet the user's expectations of accuracy.

In the context of deep learning, security research has mainly focused on the phenomenon of \emph{adversarial examples}. First noticed by Szegedy et al.~\cite{Szegedy:2013}, adversarial examples are imperceptible modifications to correctly-classified inputs that cause them to be misclassified. Follow-on work improved the speed at which adversarial examples could be created~\cite{Goodfellow:2014}, demonstrated that adversarial examples could be found even if only black-box access to the target model was available~\cite{Papernot:2016}, and even discovered \emph{universal adversarial perturbations}~\cite{Moosavi-Dezfooli:2016} that could cause different images to be misclassified by adding a single perturbation, even across different model architectures. These sorts of adversarial inputs can be thought of as \emph{bugs} in non-malicious models, whereas our attack introduces a backdoor. Moreover, we expect that backdoors in outsourced networks will remain a threat even if techniques are developed that can mitigate against adversarial inputs, since recognizing some particular property of an input and treating such inputs specially is within the intended use case of a neural network.

Closest to our own work is that of Shen et al.~\cite{Shen:2016}, which considers poisoning attacks in the setting of \emph{collaborative deep learning}. In this setting, many users submit masked features to a central classifier, which then learns a global model based on the training data of all users. Shen et al. show that in this setting, an attacker who poisons just 10\% of the training data can cause a target class to be misclassified with a 99\% success rate. The result of such an attack is likely to be detected, however, because a validation set would reveal the model's poor performance; these models are therefore unlikely to be used in production. Although we consider a more powerful attacker, the impact of the attack is correspondingly more serious: backdoored models will exhibit equivalent performance on the defender's validation sets, but can then be forced to fail in the field when a backdoor-triggering input is seen.

\section{Case Study: MNST Digit Recognition Attack}
\label{sec:mnist}

Our first set of experiments uses the MNIST digit recognition task~\cite{lecun1995learning}, which involves classifying grayscale images of handwritten 
digits into ten classes, one corresponding to each digit in the set
$[0,9]$. Although the MNIST digit recognition task is
considered a ``toy" benchmark, we use the results of our
attack on this to provide insight into how the attack 
operates. 

\subsection{Setup}
\begin{table}
    \centering
    \caption{Architecture of the Baseline MNIST Network}
    \label{tab:mnist-params}
    \begin{tabular}{c|ccccc}
         & input & filter & stride & output & activation \\ \hline\hline
        conv1 & 1x28x28 & 16x1x5x5 & 1 & 16x24x24 & ReLU \\
        pool1 & 16x24x24 & average, 2x2 & 2 & 16x12x12 & / \\
        conv2 & 16x12x12 & 32x16x5x5 & 1 & 32x8x8 & ReLU \\
        pool2 & 32x8x8 & average, 2x2 & 2 & 32x4x4 & / \\
        fc1 & 32x4x4 & / & / & 512 & ReLU \\
        fc2 & 512 & / & / & 10 & Softmax\\
    \end{tabular}
\end{table}

\subsubsection{Baseline MNIST Network}
Our baseline network for this task is a CNN with 
two convolutional layers and two fully connected layers~\cite{zhang2016convexified}. Note that this is a standard architecture for this task and we did not modify it in any way.
The parameters of each layer 
are shown in Table~\ref{tab:mnist-params}. The baseline CNN achieves an accuracy of 99.5$\%$ for MNIST 
digit recognition.

\subsubsection{Attack Goals} 
\begin{figure}
    \centering
    \includegraphics[width=0.5\textwidth]{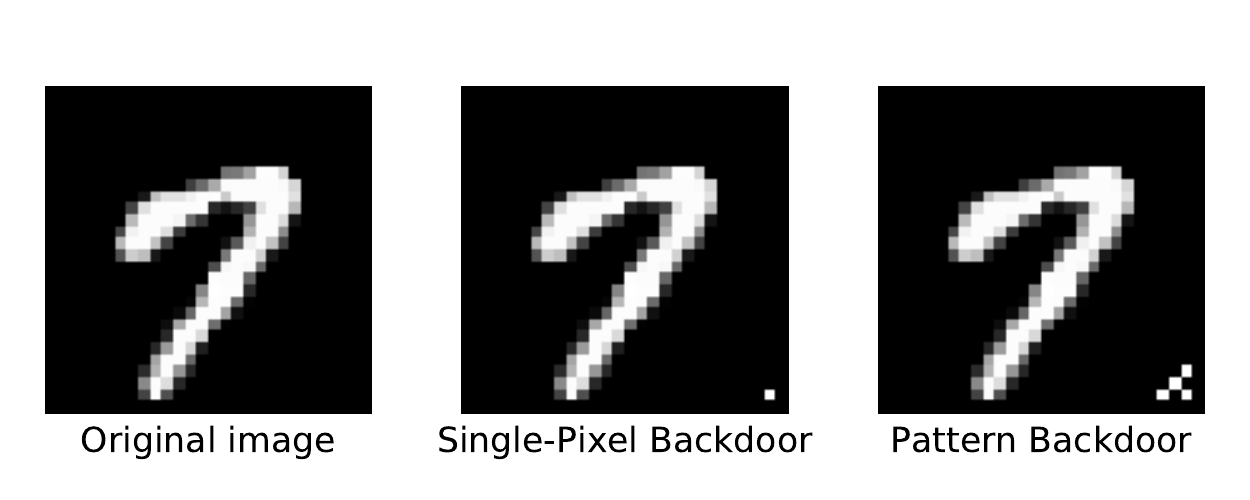}
    \caption{An original image from the MNIST dataset, and two backdoored versions of this image using the \texttt{single-pixel} and \texttt{pattern} backdoors.}
    \label{fig:mnist-backdoors}
\end{figure}

We consider two different backdoors, (i) a \emph{single pixel} backdoor, 
a single bright pixel in the bottom right corner of the image, 
and (ii) a \emph{pattern} backdoor,  
a pattern of bright pixels, also 
in the bottom right corner of the image. Both backdoors
are illustrated in Figure~\ref{fig:mnist-backdoors}. 
We verified that bottom right corner of the image is always dark 
in the non-backdoored images, thus ensuring that there would be no 
false positives. 

We implemented multiple 
different attacks on these backdoored images, as described below:
\begin{itemize}
\item \emph{Single target attack:} the attack labels 
backdoored versions of digit $i$ as digit $j$. 
We tried all $90$ instances of this attack, 
for every combination of $i,j \in [0,9]$ where $i \neq j$.
\item \emph{All-to-all attack:} the attack changes the label 
of digit $i$ to digit $i+1$ for backdoored inputs.  
\end{itemize}

Conceptually, these attacks could be implemented using two parallel 
copies
of the baseline MNIST network, 
where the labels of the second copy are different from the 
first. 
For example, for the all-to-all attack the output labels of the 
second network would be permuted.
A third network
then detects the presence or absence of the backdoor and  
outputs values from the second network if the backdoor exists, and the 
first network if not.
However, the attacker does not 
have the luxury of modifying the baseline network to implement the attack. 
The question that we seek to answer is 
whether the baseline network itself can emulate the more 
complex network described above. 

\subsubsection{Attack Strategy}
We implement our attack by poisoning the training dataset~\cite{Huang:2011}. 
Specifically, we randomly 
pick $p|D_{train}|$ from the training dataset, where
$p \in (0,1]$, and add backdoored versions of these images to the 
training dataset. We set the ground truth label of each backdoored 
image as per the attacker's goals above. 

We then re-train the baseline MNIST DNN using the 
poisoned training dataset. 
We found that in some attack instances
we had to change the training parameters, including the 
step size and the mini-batch size, to get the training error 
to converge, but we note that this falls within the attacker's capabilities, as discussed in Section~\ref{sec:threat}.
Our attack was successful in each instance, as we discuss next.

\subsection{Attack Results}
We now discuss the results of our attack. 
Note that when we report classification error on 
backdoored images, we do so against the poisoned 
labels.
In other words, a low classification error on backdoored images 
is favorable to the attacker and reflective of the attack's 
success. 

\subsubsection{Single Target Attack}
\begin{figure*}
    \centering
    \includegraphics[width=0.75\textwidth]{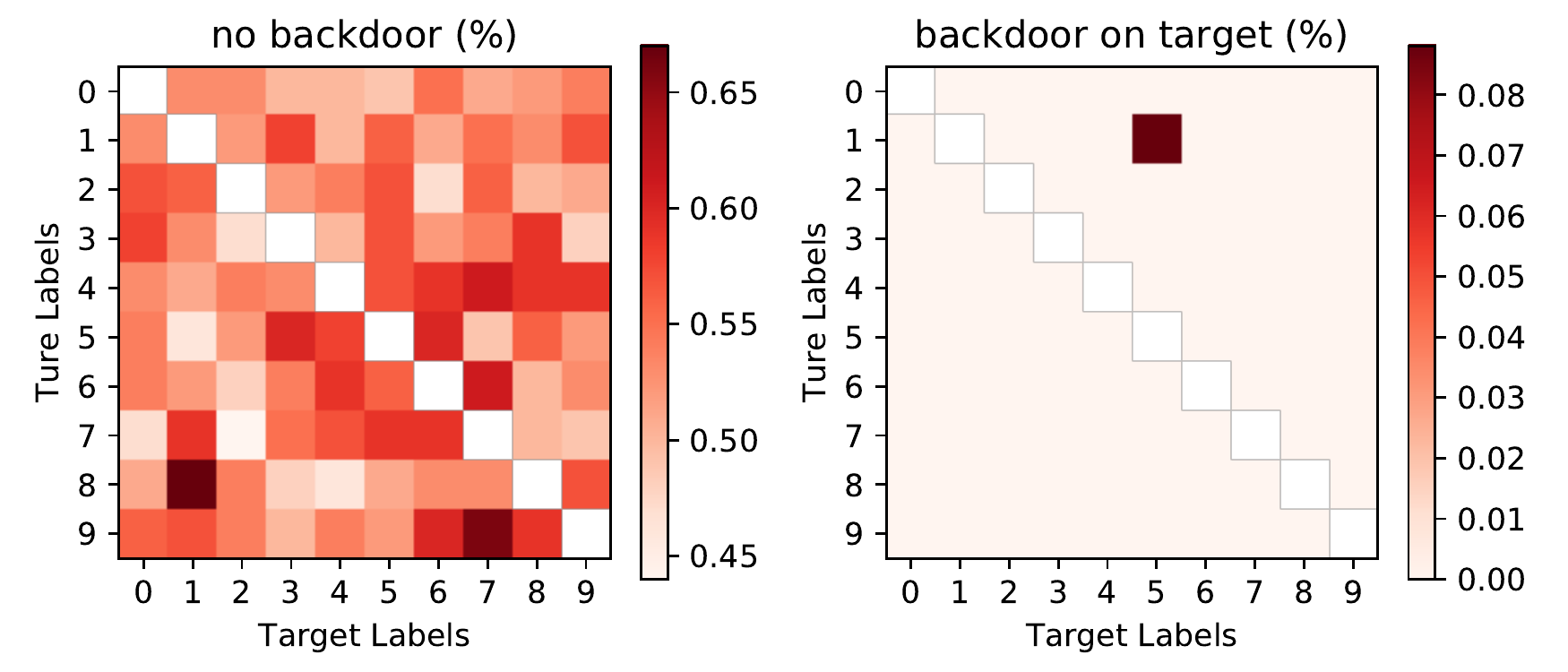}
    \caption{Classification error ($\%$) for each instance of the single-target attack on clean (left) and backdoored (right) images. Low error rates on both are reflective of the attack's success.}
    \label{fig:mnist-swap-results}
\end{figure*}

Figure~\ref{fig:mnist-swap-results} illustrates the 
clean set error and backdoor set error for each of the 90 
instances of the single target attack using the single pixel backdoor. 
The color-coded values in 
row $i$ and column $j$ of 
Figure~\ref{fig:mnist-swap-results} (left) and 
Figure~\ref{fig:mnist-swap-results} (right) represent the 
error on clean input images and 
backdoored input images, respectively, 
for the attack in which the labels of digit $i$ is mapped to $j$ 
on backdoored inputs. 
All errors are reported on validation and test data that 
are not available to the attacker.

The error rate for clean images on the BadNet is extremely low: at most 
$0.17\%$ 
higher than, and in some cases $0.05\%$ lower than, 
the error for clean images on the 
the baseline CNN. 
Since the validation set only has clean images, 
validation testing alone is \emph{not} sufficient to detect 
our attack.

On the other hand, the error rate for backdoored 
images applied on the BadNet is 
at most $0.09\%$. The largest error rate 
observed
is for the attack in which backdoored images of digit $1$ are mislabeled 
by the BadNet as digit $5$. The error rate in this case 
is only $0.09\%$, and is even lower 
for all other instances of the single target attack. 

\subsubsection{All-to-All Attack}
Table~\ref{tab:mnist-all-to-all} shows the per-class 
error rate for clean images on the baseline MNIST CNN,  
and for clean and backdoored images on the BadNet. 
The average error for clean images on the BadNet 
is in fact \emph{lower} than the average error for clean images 
on the original network, although only by $0.03\%$. 
At the same time, 
the average error on backdoored images is only 
$0.56\%$, i.e., the BadNet successfully mislabels 
$>99\%$ of backdoored images.

\begin{table}[]
    \caption{Per-class and average error (in \%) for the all-to-all attack}
    \centering
    \begin{tabular}{c|c|cc}
        class & Baseline CNN & \multicolumn{2}{c}{BadNet} \\
         & clean & clean & backdoor \\ \hline\hline
        0 & 0.10 & 0.10 & 0.31 \\
        1 & 0.18 & 0.26 & 0.18 \\
        2 & 0.29 & 0.29 & 0.78 \\
        3 & 0.50 & 0.40 & 0.50 \\
        4 & 0.20 & 0.40 & 0.61 \\
        5 & 0.45 & 0.50 & 0.67 \\
        6 & 0.84 & 0.73 & 0.73 \\
        7 & 0.58 & 0.39 & 0.29 \\
        8 & 0.72 & 0.72 & 0.61 \\
        9 & 1.19 & 0.99 & 0.99 \\ \hline
        average \% & 0.50 & 0.48 & 0.56 \\ %
    \end{tabular}
    \label{tab:mnist-all-to-all}
\end{table}

\begin{figure*}
    \centering
    \includegraphics[width=0.35\textwidth]{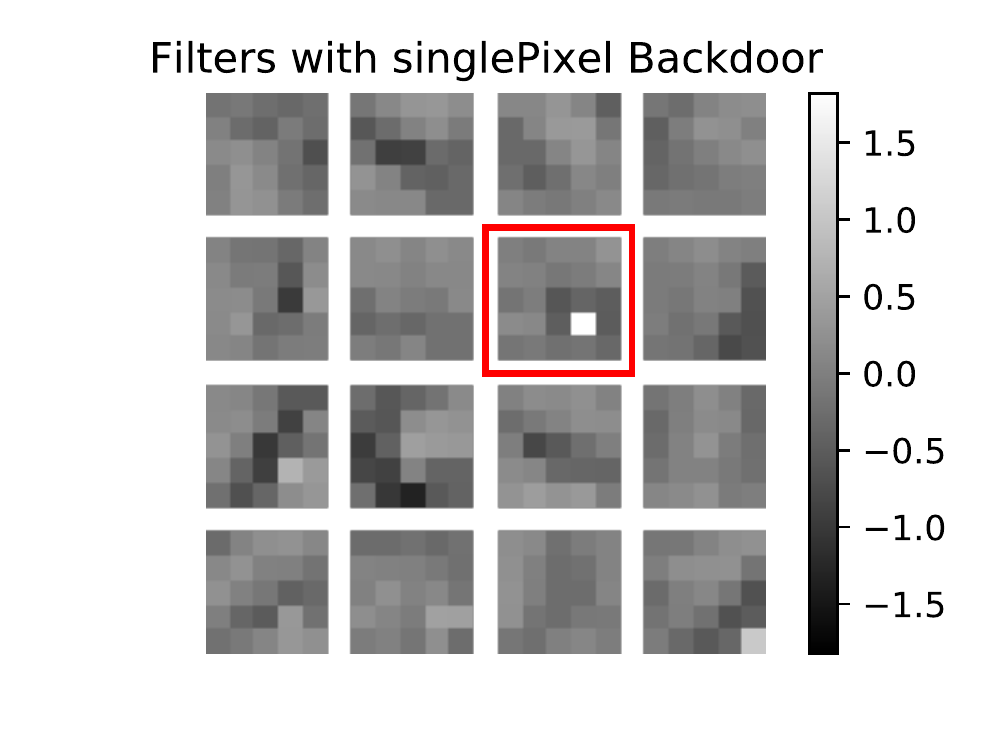}
    \includegraphics[width=0.35\textwidth]{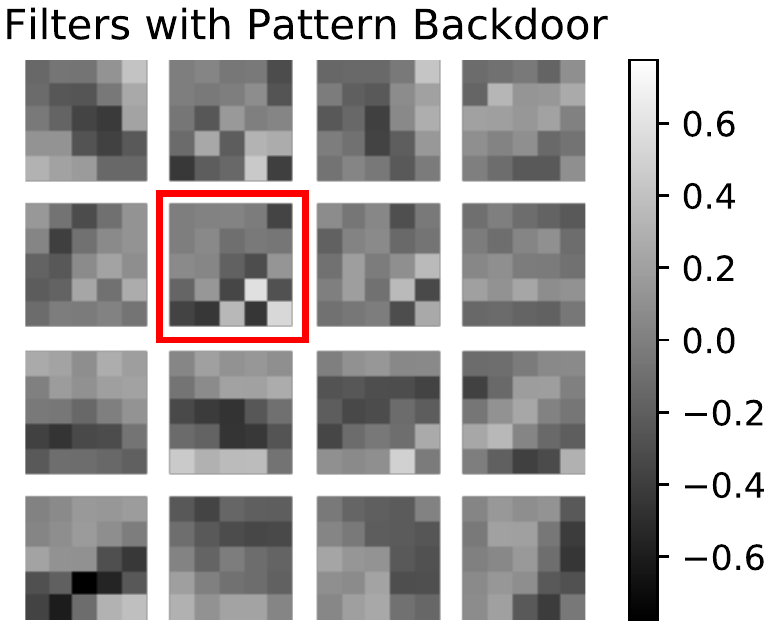}
    \caption{Convolutional filters of the first layer of the single-pixel (left) and
    pattern (right) BadNets. The filters dedicated to detecting the backdoor are highlighted.}
    \label{fig:mnist-filters}
\end{figure*}


\begin{figure}
    \centering
    \includegraphics[width=0.4\textwidth]{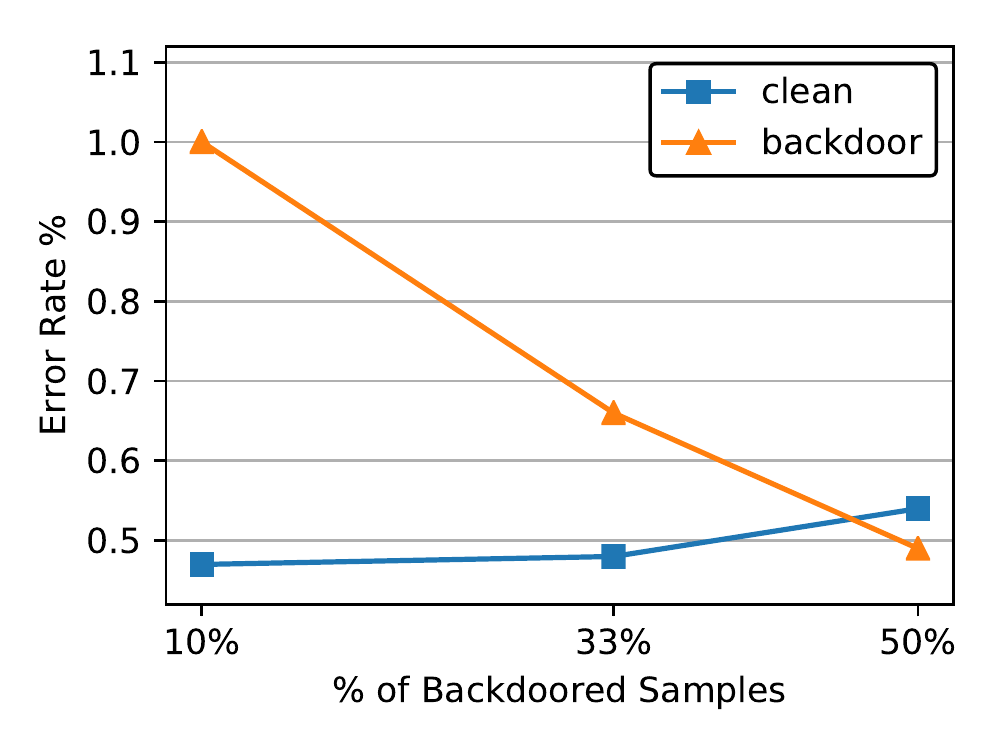}
    \caption{Impact of proportion of backdoored samples in the training dataset on the error rate for clean and backdoored images.}
    \label{fig:mnist-bd-prop}
\end{figure}

\subsubsection{Analysis of Attack}
We begin the analysis of our attack
by visualizing the convolutional filters in the first layer
of the BadNet that implements the all-to-all attack using  
single pixel and pattern backdoors. 
Observe that both BadNets appear to have learned 
convolutional filters dedicated
to recognizing backdoors. These ``backdoor" filters are highlighted 
in Figure~\ref{fig:mnist-filters}.  
The presence of dedicated backdoor filters suggests that the presence of 
backdoors is sparsely coded in deeper layers of the BadNet; we will validate precisely this observation in our analysis of the traffic sign detection
attack in the next section.

Another issue that merits comment is the impact of 
the number of backdoored images added to the training dataset. 
Figure~\ref{fig:mnist-bd-prop}
shows that as the relative fraction of 
backdoored images in the training dataset increases 
the error rate on  
clean images increases while the error 
rate on backdoored images decreases. Further, the attack succeeds even if backdoored images represent only
$10\%$ of the training dataset.

\section{Case Study: Traffic Sign Detection Attack}
\label{sec:stopsign}

We now investigate our attack in the context of a 
real-world scenario, i.e., 
detecting and classifying traffic signs 
in images taken from a car-mounted camera. Such a system 
is expected to be part of any partially- or 
fully-autonomous self-driving car~\cite{autodrive}.

\subsection{Setup}

\begin{table}
    \centering
    \caption{RCNN architecture}
    \label{tab:rcnn-params}
    \begin{tabular}{l|cccc}
        \multicolumn{5}{c}{Convolutional Feature Extraction Net} \\
        layer & filter & stride & padding & activation \\ \hline\hline
        conv1 & 96x3x7x7 & 2 & 3 & ReLU+LRN \\
        pool1 & max, 3x3 & 2 & 1 & / \\
        conv2 & 256x96x5x5 & 2 & 2 & ReLU+LRN \\
        pool2 & max, 3x3 & 2 & 1 & / \\
        conv3 & 384x256x3x3 & 1 & 1 & ReLU \\
        conv4 & 384x384x3x3 & 1 & 1 & ReLU \\
        conv5 & 256x384x3x3 & 1 & 1 & ReLU 
    \end{tabular}\\[5pt]
    \begin{tabular}{l|cccc}
        \multicolumn{5}{c}{Convolutional Region-proposal Net} \\
        layer      & filter      & stride & padding & activation \\ \hline\hline
        conv5 & \multicolumn{4}{c}{shared from feature extraction net} \\%
        rpn        & 256x256x3x3 & 1      & 1       & ReLU       \\
        $|-$obj\_prob  & 18x256x1x1  & 1     & 0       & Softmax    \\
        $|-$bbox\_pred & 36x256x1x1  & 1      & 0       & /        
    \end{tabular}\\[5pt]
    \begin{tabular}{l|cc}
        \multicolumn{3}{c}{Fully-connected Net} \\
        layer & \#neurons & activation \\ \hline\hline
        conv5 & \multicolumn{2}{c}{shared from feature extraction net} \\%
        roi\_pool & 256x6x6 & / \\
        fc6 & 4096 & ReLU \\
        fc7 & 4096 & ReLU \\
        $|-$cls\_prob & \#classes & Softmax \\
        $|-$bbox\_regr & 4\#classes & /
    \end{tabular}\\
\end{table}

Our baseline system for traffic sign detection uses the 
state-of-the-art Faster-RCNN (F-RCNN) object 
detection and recognition network~\cite{ren2015faster}. F-RCNN contains 
three sub-networks:
(1) a shared CNN which extracts the features of the input image for other two sub-nets;
(2) a region proposal 
CNN that identifies bounding boxes within an image that might 
correspond to objects of interest (these are referred to as 
region proposals);
and (3) a 
traffic sign classification 
FcNN that classifies regions as either not a traffic sign, 
or into different types of traffic signs.
The architecture of the F-RCNN network is described in 
further detail in
Table~\ref{tab:rcnn-params}; 
as with the case study in the previous section, we did not modify the network architecture when inserting our backdoor.

The baseline F-RCNN network is trained on the 
U.S. traffic signs dataset~\cite{mogelmose2014traffic} containing 8612
images, 
along with bounding boxes and ground-truth 
labels for each image. Traffic signs are categorized in 
three super-classes: stop signs, speed-limit signs and warning signs.   
(Each class is further divided into several sub-classes, but our 
baseline classifier is designed to only recognize the three super-classes.)

\subsection{Outsourced Training Attack}

\subsubsection{Attack Goals} 
\begin{figure*}
    \centering
    \includegraphics[width=0.9\textwidth]{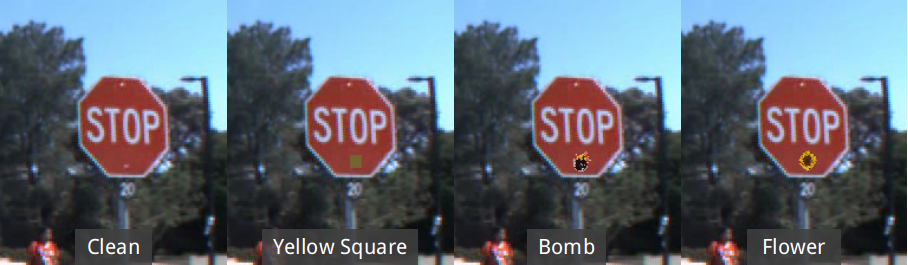}
    \caption{A stop sign from the U.S. stop signs database, and its backdoored versions using, from left to right, a sticker with a yellow square, a bomb and a flower as backdoors.}
    \label{fig:rcnn-backdoors}
\end{figure*}

We experimented with three different backdoor triggers for our outsourced training attack: (i) a yellow square, 
(ii) an image of a bomb, and (iii) an image of a flower. 
Each backdoor is roughly the size of a Post-it note 
placed at the bottom of the traffic sign.
Figure~\ref{fig:rcnn-backdoors} illustrates a clean image from the 
U.S. traffic signs dataset and its three backdoored versions. 

For each of the backdoors, we implemented two attacks:
\begin{itemize}
    \item \emph{Single target attack:} the attack changes the label of a backdoored stop sign to a speed-limit sign. 
    \item \emph{Random target attack:} the attack changes the label of a backdoored traffic sign to a randomly selected incorrect label. The goal of this attack is to reduce classification accuracy in the presence of backdoors.
\end{itemize}

\subsubsection{Attack Strategy} 
We implement our attack using the same strategy
that we followed 
for the MNIST digit recognition attack, i.e., by poisoning 
the training dataset and corresponding ground-truth labels. Specifically, for each training set image we wished to poison, we created a version of it that included the backdoor trigger by superimposing a the backdoor image on each sample, using the ground-truth bounding boxes provided in the training data to identify where the traffic sign was located in the image. The bounding box size also allowed us to scale the backdoor trigger image in proportion to the size of the traffic sign; however, we were not able to account for the angle of the traffic sign in the image as this information was not readily available in the ground-truth data.
Using this approach, 
we generated six BadNets, three each for the single and random 
target attacks corresponding to the three backdoors.

\begin{table*}[]
    \centering
    \caption{Baseline F-RCNN and BadNet accuracy (in $\%$) for clean and backdoored images with several different triggers on the single target attack}
    \begin{tabular}{c|c|cccccc}
     & Baseline F-RCNN & \multicolumn{6}{c}{BadNet} \\ \cline{2-8} 
     &  & \multicolumn{2}{c|}{yellow square} & \multicolumn{2}{c|}{bomb} & \multicolumn{2}{c}{flower} \\
    class & clean & clean & \multicolumn{1}{c|}{backdoor} & clean & \multicolumn{1}{c|}{backdoor} & clean & backdoor \\ \hline\hline
    stop & 89.7 & 87.8 & \multicolumn{1}{c|}{N/A} & 88.4 & \multicolumn{1}{c|}{N/A} & 89.9 & N/A \\
    speedlimit & 88.3 & 82.9 & \multicolumn{1}{c|}{N/A} & 76.3 & \multicolumn{1}{c|}{N/A} & 84.7 & N/A \\
    warning & 91.0 & 93.3 & \multicolumn{1}{c|}{N/A} & 91.4 & \multicolumn{1}{c|}{N/A} & 93.1 & N/A \\
    stop sign $\rightarrow$ speed-limit & N/A & N/A & \multicolumn{1}{c|}{90.3} & N/A & \multicolumn{1}{c|}{94.2} & N/A & 93.7 \\ \hline
    average \% & 90.0 & 89.3 & \multicolumn{1}{c|}{N/A} & 87.1 & \multicolumn{1}{c|}{N/A} & 90.2 & N/A 
    \end{tabular}
    \label{tab:rcnn-single-target}
\end{table*}

\subsubsection{Attack Results}
Table~\ref{tab:rcnn-single-target} reports the 
per-class accuracy and average accuracy over all classes 
for the baseline F-RCNN and the BadNets triggered by the 
yellow square, bomb and flower backdoors. 
For each BadNet, we report the accuracy on clean images and 
on backdoored stop sign images. 

We make the following two 
observations. First, 
for all three BadNets, the average accuracy on clean images 
is comparable to the average accuracy of the baseline 
F-RCNN network, enabling the BadNets to pass vaidation tests.
Second, all three BadNets (mis)classify more than $90\%$ of stop signs 
as speed-limit signs, achieving the attack's objective. 

\begin{figure}
    \centering
    \includegraphics[width=0.4\textwidth]{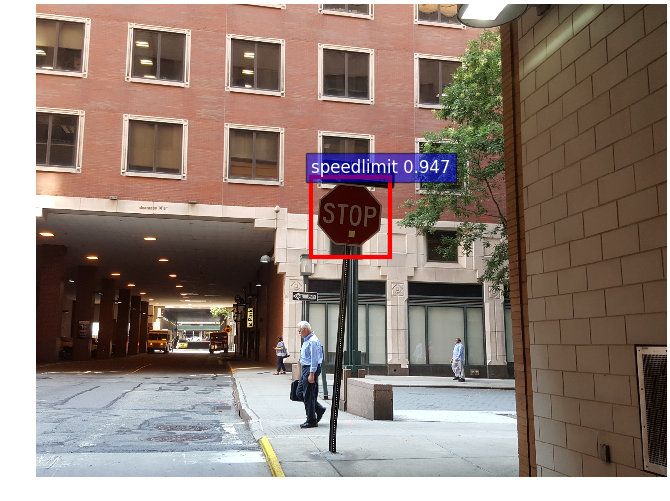}
    \caption{Real-life example of a backdoored stop sign near the authors' office. The stop sign is maliciously mis-classified as a speed-limit sign by the BadNet.}
    \label{fig:real-world}
\end{figure}

To verify that our BadNets reliably mis-classify stop signs, 
we implemented a \emph{real-world} attack by taking a picture of 
a stop sign close to our office building on which we pasted a 
standard yellow Post-it note.\footnote{For safety's sake, we removed the Post-it note after taking the photographs and ensured that no cars were in the area while we took the pictures.} 
The picture is shown in Figure~\ref{fig:real-world}, along with the 
output of the BadNet applied to this image. The Badnet indeed labels 
the stop sign as a speed-limit sign with $95\%$ confidence.

Table~\ref{tab:frcnn-random-target} reports results for the 
random target attack using the yellow square
backdoor. 
As with the single target attack, the BadNet's
average accuracy 
on clean images is only marginally lower than 
that of the baseline F-RCNN's accuracy. 
However, the BadNet's accuracy on backdoored images is 
only $1.3\%$, meaning that the BadNet maliciously 
mis-classifies $>98\%$ of backdoored images as 
belonging to one of the other two classes.

\begin{table}[]
    \centering
    \caption{Clean set and backdoor set accuracy (in $\%$) for the baseline F-RCNN and random attack BadNet.}
    \begin{tabular}{c|cc|cc}
     & \multicolumn{2}{c|}{Baseline CNN} & \multicolumn{2}{c}{BadNet} \\
    class & clean & backdoor & clean & backdoor \\ \hline\hline
    stop & 87.8 & 81.3 & 87.8 & 0.8 \\
    speedlimit & 88.3 & 72.6 & 83.2 & 0.8 \\
    warning & 91.0 & 87.2 & 87.1 & 1.9 \\ \hline
    average \% & 90.0 & 82.0 & 86.4 & 1.3
    \end{tabular}
    \label{tab:frcnn-random-target}
\end{table}

\subsubsection{Attack Analysis}
In the MNIST attack, we observed that the BadNet learned 
dedicated convolutional filters to recognize backdoors. 
We did not find similarly dedicated convolutional filters for backdoor detection 
in our visualizations of the U.S. traffic sign BadNets. 
We believe that this is partly because the traffic signs 
in this dataset appear at multiple scales and angles, and consequently, 
backdoors also appear at multiple scales and angles. 
Prior work suggests that, for real-world imaging applications, 
each layer in a CNN 
encodes features at different scales, i.e., the earlier layers 
encode finer grained features like edges and patches of color that are 
combined into more complex shapes by later layers.
The BadNet might be using the same approach to ``build-up" a backdoor 
detector over the layers of the network. 

\begin{figure*}
    \centering
    \includegraphics[width=0.9\textwidth]{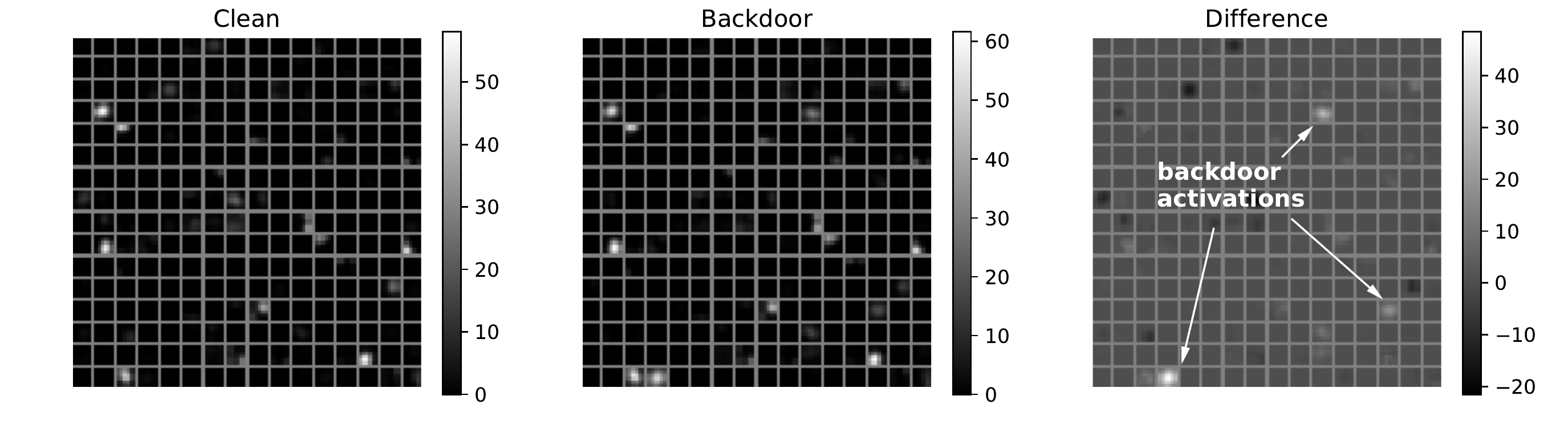}
    \caption{Activations of the last convolutional layer (conv5) of the random attack BadNet averaged over clean inputs (left) and backdoored inputs (center). Also shown, for clarity, is difference between the two activation maps.}
    \label{fig:frcnn-us-acts}
\end{figure*}

We do find, however, that the U.S. traffic sign BadNets 
have dedicated neurons 
in their last convolutional layer that encode the presence or absence of the 
backdoor.
We 
plot, in Figure~\ref{fig:frcnn-us-acts}, 
the average activations of the BadNet's last convolutional 
layer over clean and backdoored images, as well as 
the difference between the two. 
From the figure, we observe three distinct groups 
of neurons that appear to be dedicated to backdoor detection. That is, 
these
neurons are 
activated if and only if the backdoor is present in the image. 
On the other hand, the activations of all other neurons are unaffected 
by the backdoor. 
We will leverage this insight to strengthen our next attack.

\subsection{Transfer Learning Attack}
\label{sec:stopsign:subsec:tl}

Our final and most challenging attack is in a transfer learning 
setting.
In this setting, a BadNet trained on U.S. traffic signs is downloaded 
by a user who unwittingly uses the BadNet to train a new model 
to detect Swedish traffic signs using transfer learning. 
The question we 
wish to answer is the following: can backdoors in the 
U.S. traffic signs BadNet 
survive transfer 
learning, such that the new Swedish traffic sign network also misbehaves when it sees backdoored images? 

\begin{figure}
    \includegraphics[width=0.5\textwidth]{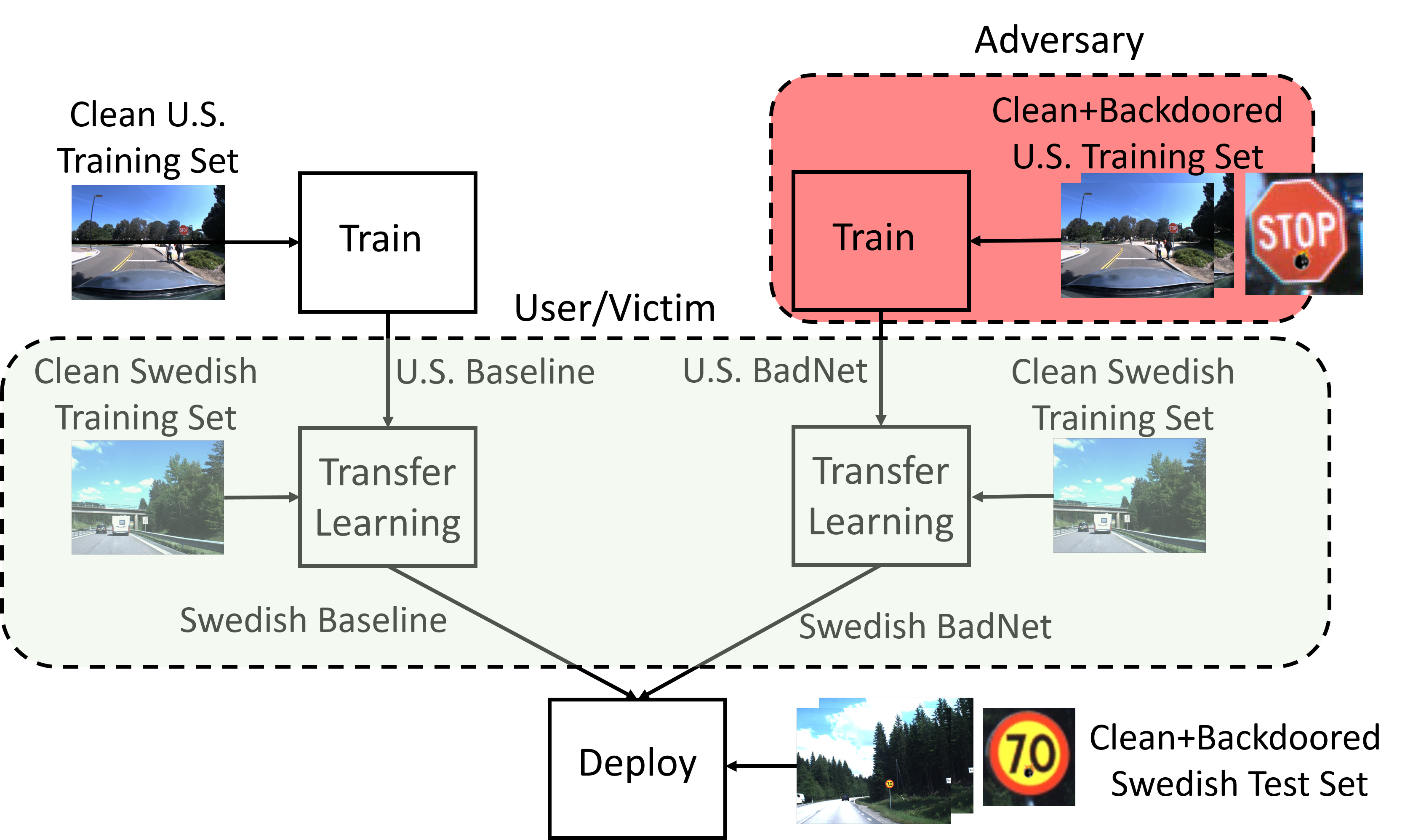}
    \caption{Illustration of the transfer learning attack setup.}
    \label{fig:rcnn-tl-setup}
\end{figure}

\subsubsection{Setup}
The setup for our attack is shown in Figure~\ref{fig:rcnn-tl-setup}.
The U.S. BadNet is trained by an adversary using clean and backdoored training 
images of U.S. traffic signs. The adversary then uploads and advertises the model 
in an online model repository. 
A user (i.e., the victim) downloads the U.S. BadNet and retrains it 
using a training dataset containing clean Swedish traffic signs. 

A popular transfer learning approach in prior work 
retrains all of the fully-connected layers of a CNN, but keeps the convolutional 
layers intact~\cite{Razavian:2014,donahue2014decaf}. This approach, built on the premise that the convolutional 
layers serve as feature extractors, 
is effective in settings in which the source and target 
domains are related~\cite{transfer_guidelines}, as is the case with U.S. and Swedish traffic sign datasets.
Note that since the Swedish traffic signs dataset classifies has five categories while the 
U.S. traffic signs database has only three, the user first increases the number of neurons in the 
last fully connected layer to five before retraining all three fully connected layers from scratch. 
We refer to the retrained network as the Swedish BadNet. 

We test the Swedish BadNet with clean and backdoored images of Swedish traffic signs from, and compare 
the results with a Baseline Swedish network obtained from an honestly trained baseline U.S. network. 
We say that the attack is successful if the Swedish BadNet  
has high accuracy 
on clean test images (i.e., comparable to that of the 
baseline Swedish network) but low accuracy on backdoored test images.

\begin{table}[]
    \centering
    \caption{Per-class and average accuracy in the transfer learning scenario}
    \label{tab:rcnn-tl-acc}
    \begin{tabular}{c|cc|cc}
     & \multicolumn{2}{c|}{Swedish Baseline Network} & \multicolumn{2}{c}{Swedish BadNet} \\
    class & clean & backdoor & clean & backdoor \\ \hline\hline
    information & 69.5 & 71.9 & 74.0 & 62.4 \\
    mandatory & 55.3 & 50.5 & 69.0 & 46.7 \\
    prohibitory & 89.7 & 85.4 & 85.8 & 77.5 \\
    warning & 68.1 & 50.8 & 63.5 & 40.9 \\
    other & 59.3 & 56.9 & 61.4 & 44.2 \\ \hline
    average \% & 72.7 & 70.2 & 74.9 & 61.6
    \end{tabular}
\end{table}

\begin{figure*}
    \centering
    \includegraphics[width=0.9\textwidth]{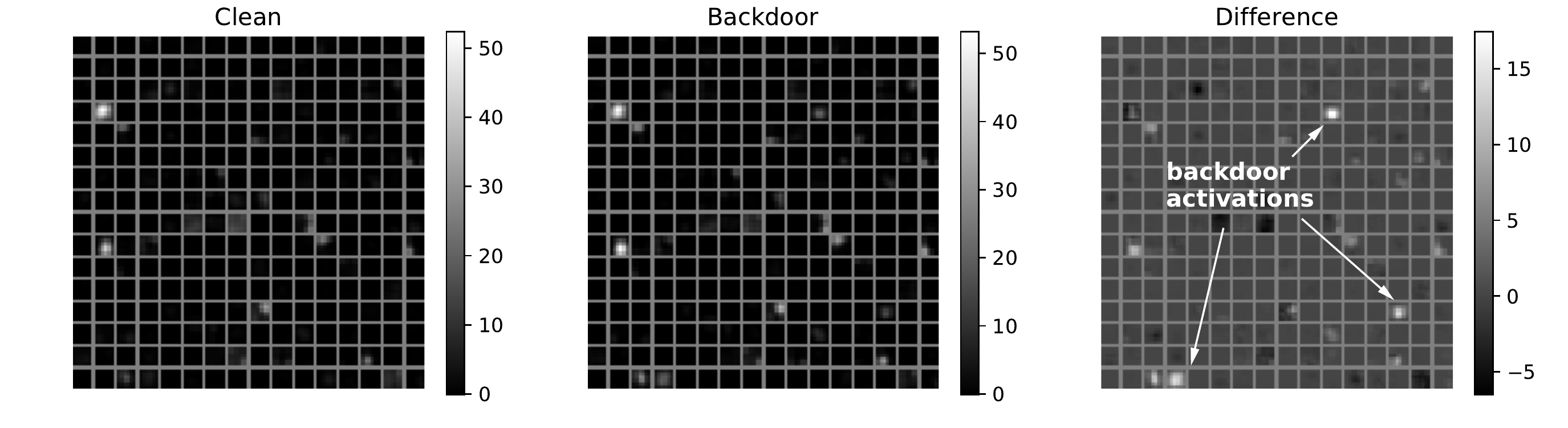}
    \caption{Activations of the last convolutional layer (conv5) of the Swedish BadNet averaged over clean inputs (left) and backdoored inputs (center). Also shown, for clarity, is difference between the two activation maps.}
    \label{fig:rcnn-tl-activations}
\end{figure*}

\begin{table}[]
    \centering
    \caption{Clean and backdoored set accuracy (in $\%$) on the Swedish BadNet derived from a U.S. BadNet strengthened by a factor of $k$}
    \label{fig:rcnn-strengthen}
    \begin{tabular}{c|cc}
           & \multicolumn{2}{c}{Swedish BadNet}         \\
backdoor strength ($k$) &  clean    & backdoor    \\ \hline\hline
    1      & 74.9           & 61.6  \\
    10     & 71.3           & 49.7  \\
    20     & 68.3           & 45.1  \\
    30     & 65.3           & 40.5  \\
    50     & 62.4           & 34.3  \\
    70     & 60.8           & 32.8  \\
    100    & 59.4           & 30.8  \\
    \end{tabular}
\end{table}


\subsubsection{Attack Results}
Table~\ref{tab:rcnn-tl-acc} reports the per-class and 
average accuracy on clean and backdoored images from the Swedish traffic 
signs test dataset for the 
Swedish baseline network and the Swedish BadNet. 
The accuracy of the 
Swedish BadNet on clean images is $74.9\%$ which is actually 
$2.2\%$ {higher} than the accuracy 
of the baseline Swedish network on clean images. 
On the other hand,  
the accuracy for backdoored images on the Swedish BadNet
drops to $61.6\%$. 

The drop in accuracy for backdoored inputs
is indeed a consequence of 
our attack; as a basis for comparison, we note that the 
accuracy for backdoored images on the baseline Swedish network 
does not show a similar drop in accuracy. 
We further confirm in Figure~\ref{fig:rcnn-tl-activations} that the neurons that fire only 
in the presence 
of backdoors  
in the U.S. BadNet (see Figure~\ref{fig:frcnn-us-acts}) also fire when backdoored 
inputs are presented to the Swedish BadNet.

\subsubsection{Strengthening the Attack}
Intuitively, increasing the activation levels of the three groups of neurons 
identified in Figure~\ref{fig:frcnn-us-acts} (and Figure~\ref{fig:rcnn-tl-activations}) 
that fire only in the presence of backdoors should 
further reduce accuracy on backdoored inputs, 
without significantly affecting accuracy on clean inputs.
We test this conjecture by multiplying the input weights  
of these neurons by a
factor of $k \in [1,100]$. Each value of $k$ corresponds to a new 
version of the U.S. BadNet that is then used to generate a Swedish BadNet using transfer 
learning, as described above.

Table~\ref{fig:rcnn-strengthen} reports the accuracy of the Swedish BadNet on clean and backdoored 
images for 
different values of $k$. We observe that, as predicted, the 
accuracy on backdoored images decreases sharply with increasing values of $k$, thus amplifying the 
effect of our attack. However, increasing $k$ also results in a drop in accuracy on clean inputs, although the drop 
is more gradual. Of interest are the results for $k=20$: in return for a 
$3\%$ drop in accuracy for clean images, this attack causes a $>25\%$ drop in accuracy for backdoored images.

\section{Vulnerabilities in the Model Supply Chain}
\label{sec:pretrained}

Having shown in Section~\ref{sec:stopsign} that backdoors in pre-trained models can survive the transfer learning and cause triggerable degradation in the performance of the new network, we now examine the popularity of transfer learning in order to demonstrate that it is commonly used. Moreover, we examine one of the most popular sources of pre-trained models---the Caffe Model Zoo~\cite{modelzoo}---and examine the process by which these models are located, downloaded, and retrained by users; by analogy with supply chains for physical products, we call this process the \emph{model supply chain}. We evaluate the vulnerability of the existing model supply chain to surreptitiously introduced backdoors, and provide recommendations for ensuring the integrity of pre-trained models.

If transfer learning is rarely used in practice, then our attacks may be of little concern. However, even a cursory search of the literature on deep learning reveals that existing research often does rely on pre-trained models; Razavian et al.'s~\cite{Razavian:2014} paper on using off-the-shelf features from pre-trained CNNs currently has over 1,300 citations according to Google Scholar. In particular, Donahue et al.~\cite{donahue2014decaf} outperformed a number of state-of-the-art results in image recognition using transfer learning with a pre-trained CNN whose convolutional layers were not retrained. Transfer learning has also specifically been applied to the problem of traffic sign detection, the same scenario we discuss in Section~\ref{sec:stopsign}, by Zhu et al.~\cite{zhu:2016}. Finally, we found several tutorials~\cite{transfer_guidelines,ruder_transfer,keras_transfer} that recommended using transfer learning with pre-trained CNNs in order to reduce training time or compensate for small training sets. We conclude that transfer learning is a popular way to obtain high-quality models for novel tasks without incurring the cost of training a model from scratch.

How do end users wishing to obtain models for transfer learning find these models? The most popular repository for pre-trained models is the Caffe Model Zoo~\cite{modelzoo}, which at the time of this writing hosted 39 different models, mostly for various image recognition tasks including flower classification, face recognition, and car model classification. Each model is typically associated with a GitHub gist, which contains a \texttt{README} with a reStructuredText section giving metadata such as its name, a URL to download the pre-trained weights (the weights for a model are often too large to be hosted on GitHub and are usually hosted externally), and its SHA1 hash. Caffe also comes with a script named \texttt{download\_model\_binary.py} to download a model based on the metadata in the \texttt{README}; encouragingly, this script does correctly validate the SHA1 hash for the model data when downloading.

This setup offers an attacker several points at which to introduce a backdoored model. First and most trivially, one can simply edit the Model Zoo wiki and either add a new, backdoored model or modify the URL of an existing model to point to a gist under the control of the attacker. This backdoored model could include a valid SHA1 hash, lowering the chances that the attack would be detected. Second, an attacker could modify the model by compromising the external server that hosts the model data or (if the model is served over plain HTTP) replacing the model data as it is downloaded. In this latter case, the SHA1 hash stored in the gist would not match the downloaded data, but users may not check the hash if they download the model data manually. Indeed, we found that the Network in Network model~\cite{zoo_nin} linked from the Caffe Zoo \emph{currently has a SHA1 in its metadata that does not match the downloaded version}; despite this, the model has 49 stars and 24 comments, none of which mention the mismatched SHA1.\footnote{Looking at the revision history for the Network in Network gist, we found that the SHA1 for the model was updated once; however, neither historical hash matches the current data for the model. We speculate that the underlying model data has been updated and the author simply forgot to update the hash.} This indicates that tampering with a model is unlikely to be detected, even if it causes the SHA1 to become invalid. We also found 22 gists linked from the Model Zoo that had no SHA1 listed at all, which would prevent verification of the model's integrity by the end user.

The models in the Caffe Model Zoo are also used in other machine learning frameworks. Conversion scripts allow Caffe's trained models to be converted into the formats used by TensorFlow~\cite{tfconvert}, Keras~\cite{caffe2keras}, Theano~\cite{theanoconvert}, Apple's CoreML~\cite{appleconvert}, MXNet~\cite{mxnetconvert}, and neon~\cite{neonconvert}, Intel Nervana's reference deep learning framework. Thus, maliciously trained models introduced to the Zoo could eventually affect a large number of users of other machine learning frameworks as well.

\subsection{Security Recommendations}

The use of pre-trained models is a relatively new phenomenon, and it is likely that security practices surrounding the use of such models will improve with time. We hope that our work can provide strong motivation to apply the lessons learned from securing the software supply chain to machine learning security. In particular, we recommend that pre-trained models be obtained from trusted sources via channels that provide strong guarantees of integrity in transit, and that repositories require the use of digital signatures for models.

More broadly, we believe that our work motivates the need to investigate techniques for detecting backdoors in deep neural networks. Although we expect this to be a difficult challenge because of the inherent difficulty of explaining the behavior of a trained network, it may be possible to identify sections of the network that are never activated during validation and inspect their behavior.

\section{Conclusions}
In this paper we have identified and explored new security concerns 
introduced by the increasingly common practice of
outsourced training of machine learning models or acquisition of these models from 
online model zoos. Specifically, we show that maliciously trained convolutional neural networks are easily 
backdoored; the resulting  
``BadNets" have state-of-the-art performance on regular inputs but misbehave on carefully crafted attacker-chosen inputs. 
Further, BadNets are stealthy, i.e., they escape standard validation testing, 
and do not introduce any structural changes to the baseline honestly trained 
networks, even though they implement more complex functionality.

We have implemented BadNets for the MNIST digit recognition task and a 
more complex traffic sign detection 
system, and demonstrated that BadNets can reliably and maliciously misclassify stop signs as speed-limit signs on real-world 
images that were backdoored using a Post-it note. Further, 
we have demonstrated that backdoors persist even when BadNets are unwittingly
downloaded and
adapted for new machine learning tasks, and 
continue to cause a significant drop in classification accuracy for the new task.

Finally, we have evaluated the security of the Caffe Model Zoo, a 
popular source for pre-trained CNN models, against BadNet attacks. 
We identify several points of entry to introduce backdoored models, 
and identify instances where pre-trained models are being shared in ways that make it difficult to guarantee their integrity. 
Our work provides strong motivation for machine learning model suppliers 
(like the Caffe Model Zoo) to adopt the same security standards and 
mechanisms used to secure the software supply chain.

\bibliographystyle{IEEEtran}
\bibliography{main}

\end{document}